\documentstyle[11pt,newpasp,twoside,epsfig]{article}
\def\edcomment#1{\iffalse\marginpar{\raggedright\sl#1\/}\else\relax\fi}
\reversemarginpar

\begin{document}
\title{The Fundamental Plane of Black Hole Activity and
  the Census of the Local Black Holes' Population}
\author{Andrea Merloni}
\affil{Max-Planck-Institut f\"ur Astrophysik, 
Karl-Schwarzschild-Str. 1; D-85741 Garching, Germany.}

\begin{abstract}
 Studying a sample of both strongly and weakly active
  galactic nuclei with measured masses and 5 GHz and 2-10 keV
  core luminosities, together with a few galactic black holes 
 simultaneously observed in the radio and X-ray bands, Merloni,
  Heinz, \& Di Matteo (2003) showed
  that the sources are correlated through a ``fundamental
  plane'' relationship in the three-dimensional
  ($\log L_{\rm R},\log L_{\rm X},\log M$) space. 
  Here I elaborate on how such a relationship can be used to infer
  directly mass and accretion rate of any black hole given its
  radio and X-ray fluxes, 
  complementing the information obtained from optical/UV surveys. As an
  example, I show how the local X-ray and radio luminosity functions,
  coupled with the black hole mass function derived from the
  $M-\sigma$ relation, provide us with an accretion rate
  function. We found that the typical X-ray Eddington ratio
  of an active black hole at redshift zero is about $5 \times 10^{-4}$.

\end{abstract}

\section{Introduction}

In a recent paper Merloni, Heinz, \& Di Matteo (2003) have
shown that disc and jet emission from 
active black holes of any mass, from galactic X-ray binary
sources to the most powerful quasars, are physically and
observationally correlated phenomena. 
Their main result is the discovery of a ``fundamental plane'' of
black hole activity; that is, if we define the instantaneous state of
activity of a black hole of mass $M$ (in units of solar masses), by the
radio and hard X-ray luminosity of its
compact core, and represent such an object as a point in the
three-dimensional space ($\log L_{\rm R},\log L_{\rm X},\log M$), all 
black holes (either of stellar mass or supermassive)
will lie preferentially on a plane, described by the following
equation:
\begin{equation}
\log L_{\rm R}=(0.60^{+0.11}_{-0.11}) \log L_{\rm X}
+(0.78^{+0.11}_{-0.09}) \log M + 7.33^{+4.05}_{-4.07}.
\label{eq:fp}
\end{equation}

There are two main reasons why such a long sought-after correlation
has been discovered just now. The first is the importance of having
large numbers of accurately measured black hole masses, which only became
available in the {\it HST} era thanks to the exquisite spatial
resolution needed for this kind of dynamical studies (see
e.g. Magorrian et al. 1998). Moreover, the tight empirical correlation
between black hole masses and central velocity dispersion
of the host's bulge 
now allows to infer BH masses from larger scale
galactic properties, greatly increasing  the number of reliable
mass estimates available, at least in the local universe.

The second crucial factor is
 the identification of the hard (2-10 keV)
X-ray spectral range as the best suited to unveil accretion activity,
because of the little importance of absorption in that band
(barred Compton thick sources, of course). However, the search for hard
X-ray emission in all but the brightest galactic nuclei had to wait
until the launch of the {\it Chandra} satellite before it could be
carried on in a systematic fashion (see e.g. Ho et al 2001). 

Although currently marred by a large intrinsic scatter (see the
discussion in Merloni at al. 2003), the fundamental plane relationship
opens the way to a number of potentially important applications in the
study of large samples of black holes, not least because the physics
behind it is reasonably well understood 
(Begelman, Blandford, \& Rees 1984; Falcke \& Biermann 1995; Merloni et al. 2003).
Here I would like to point out a simple way of using the fundamental
plane relation as a (gross) mass {\it and accretion rate} estimator.
Coupling the two well determined local radio and X-ray
luminosity functions of AGN through equation (1), it is possible 
to infer the main properties (mass and accretion rate) of the entire
population of supermassive black holes (SMBH) in the local universe.
Within the conference perspective, the goal of such an exercise is to
highlight the importance of a multiwavelength approach to SMBH
population studies. If large optical/UV surveys as SDSS 
are still crucial for determining BH masses (either directly or via
the $M- \sigma$ relation), X-rays and radio data provide 
important clues on BH activity and physical state.

\section{Measuring Masses and Accretion Rates}

Equation (1) can be inverted to  relate BH masses 
to observed nuclear radio and X-ray fluxes and the
distance $D$ (in Mpc):
\begin{equation}
 \log M \simeq 16.3 + \log D + 1.28 (\log F_{\rm R} - 0.60 \log F_{\rm
  X}) \pm 1.06.
\end{equation}
This is an entirely empirical relation, and as such is model independent.
On the other hand, it is also possible to relate the observed X-ray and
radio luminosities to the second fundamental physical parameter that
characterizes any active black hole: its accretion
rate in units of Eddington luminosity, 
$\dot m \equiv \epsilon \dot M c^2 / L_{\rm Edd}$ ($\epsilon$ is
the radiative efficiency). Such an inversion, however, is model dependent.
The reader is referred to Merloni et al. (2003) for a thorough
discussion on how to constrain disc-jet models on the basis of the
observed correlation coefficients. There they showed that the 
fundamental plane is consistent with the most general
theoretical relation between radio emission, mass and accretion
rate expected from synchrotron emitting jets (regardless of their
detailed geometrical and kinematical properties), provided that the
X-ray emitting flow is radiatively inefficient. In this case, $L_{\rm
  X} \propto \dot m^{2.3}$, and the radio
luminosity satisfies:
\begin{equation}
\label{eq:lr_bigmdot}
L_{\rm R} \propto \dot{m}^{1.38} M^{1.38} = \dot{M}^{1.38},
\end{equation}
i.e., $L_{\rm R}$ scales with the {\em physical accretion rate} only.

\begin{figure}
\plotfiddle{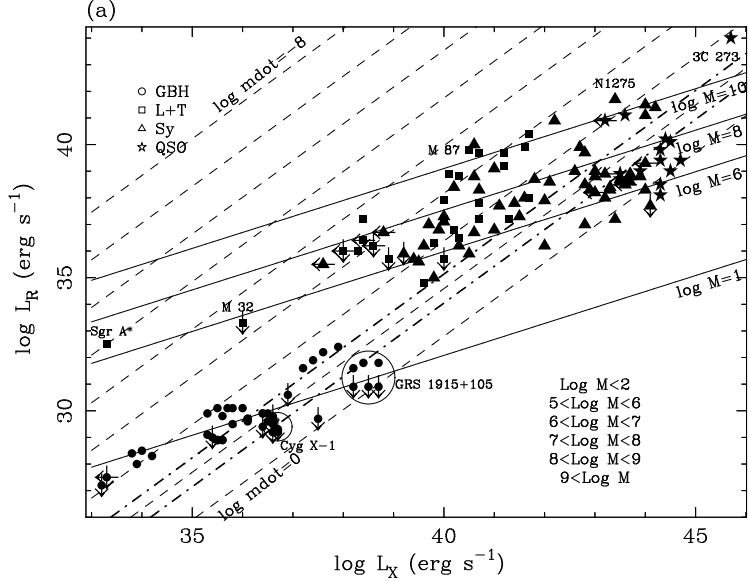}{6.5cm}{0}{60}{60}{-160}{-15}
\caption{The observed radio and X-ray core luminosity for the sample
  studied in Merloni et al. (2003). Superimposed are the lines of
  constant mass and accretion rate (in units of the Eddington
  luminosity, assuming a 10\% efficiency). Dot-dashed lines
  bound the range of accretion rates where a mode change might
  take place between radiatively efficient and inefficient flows.}
\end{figure}

Taken together, eqs. (2) and (3) can then be used to build a one to one map
of the $L_{\rm R} - L_{\rm X}$ plane onto the $M - \dot m$ one, as
shown in Figure 1.

\section{The Census of Local Black Holes' Accretion Rate}
The black hole mass function could be easily recovered if we knew the
{\it conditional} radio/X-ray luminosity function $\Phi_{\rm C}(\log L_{\rm
  X}|\log L_{\rm R})$, namely the number
of sources per unit volume with a given X-ray and radio luminosity:
\begin{equation}
\phi_{M} (\log M) \Delta \log M = \int_{(\log M,}\int_{\Delta \log M)} 
\Phi_{\rm C}(\log L_{\rm
  X}|\log L_{\rm R}) d \log L_{\rm X} d\log L_{\rm R},  
\end{equation}  
where the double integral is performed over the range of radio and
X-ray luminosities such that the logarithm of the 
mass derived from eq. (2) lies between
$\log M$ and $\log M + \Delta \log M$. An analogous expression could
be written down for the accretion rate function $\phi_{\dot m}$, in
which equation (3) can be used to determine the area over which the 
integral must be performed. 

While the $\Phi_{\rm C}$ is not known, we do know 
the two separately determined X-ray and radio luminosity
functions. Here I adopt for the 2-10 keV one,
$\phi_{\rm X}(\log L_{\rm X})$, that of Ueda et al. (2003), 
while for the 5 GHz one, $\phi_{\rm R}(\log L_{\rm R})$, that
 determined at 1.4 GHz by Sadler et al (2002), assuming a radio
 spectral index of $-0.7$. Then I assume that it is possible to
factorize the unknown conditional luminosity function as
\begin{equation}
\Phi_{\rm C}(\log L_{\rm X}|\log L_{\rm R})=\phi_{\rm X}(\log L_{\rm X})\phi_{\rm
R}(\log L_{\rm R}) f(\log L_{\rm X},\log L_{\rm R}),
\end{equation}
where $f$ is a ``matching'' function
that counts the number of object with X-ray and radio luminosities
given by $L_{\rm X}$ and $L_{\rm R}$, respectively. In such a way our
ignorance about the true conditional luminosity function is
enclosed in the matching function $f$. 
A straightforward way around this obstacle is to use the local
black hole mass function as determined from the $M - \sigma$
relationship (see e.g. Aller \& Richstone 2002) to invert
eq. (4), find the matching function $f$ and use it to calculate the
accretion rate function (more details in Merloni 2004, in
preparation). 
In Figure 2, I show the
result of such a calculation in the form of
 accretion rate and mass  functions of
local SMBH.  As can be seen
in Fig. 2a, the population of local black holes is dominated by
sources shining, in the X-ray band, below 10$^{-3}$ of the Eddington
luminosity (assuming a 10\% efficiency). This is in agreement 
what the average value found using the X-ray luminosity function of
Seyfert 1 galaxies alone (Page 2001), or with theoretical estimates
based on semianalitic Press-Schechter calculation of black hole growth
in CDM universes (Haiman \& Menou 2000).

\begin{figure}
\plottwo{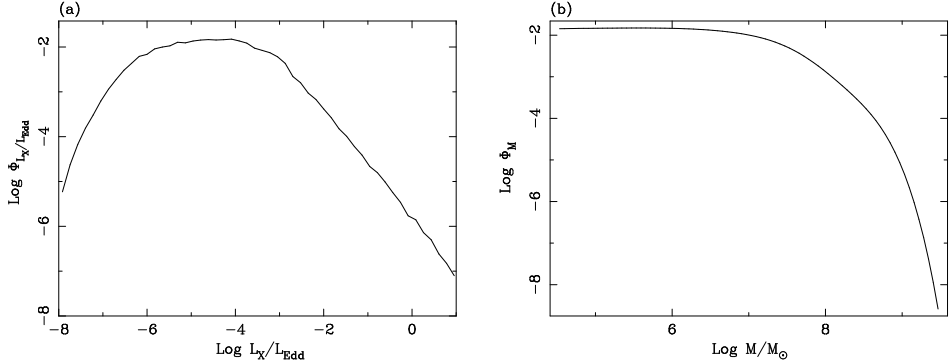}{fig2b_merloni.eps}
\caption{Local accretion rate in Eddington units (panel a, estimated from the observed
  X-ray luminosity) and mass (panel b) functions
  for supermassive black holes in the nuclei of galaxies. The
  $y$ axes show the number of sources per unit logarithmic interval
  per cubic Mpc.}
\end{figure}


\begin{references}

\reference Aller, M. C., \& Richstone, D., 2002, \aj, 124, 3035 

\reference Begelman, M. C., Blandford, R. D., \& Rees, M. J., 1984,
Rev. Mod. Phys., 56, 255

\reference Falcke, H., \& Biermann, P. L., 1995, \aap, 293, 665

\reference Haiman, Z., \& Menou, K., 2000, \apj, 531, 42

\reference Ho, L. C., et al, 2001, \apj, 549, L51

\reference Magorrian, J. et al., 1998, \aj, 115, 2285

\reference Merloni, A., Heinz, S., \& Di Matteo, T., 2003, \mnras,
345, 1057

\reference Page, M. J., 2001, \mnras, 328, 925

\reference Sadler, E. M., et al, 2002, \mnras, 329, 227

\reference Ueda, Y., Akiyama, M., Ohta, K., \& Miyaji, T., 2003, \apj,
598, 886. 

\end{references}
\end{document}